# Digital Twins of Mechanically Ventilated Preterm Neonates with Respiratory Distress Syndrome

Sina Saffaran PhD[1], Tng Chang Kwok MD[2,3], Don Sharkey MD[2,3], and Declan G. Bates PhD [*,1]

1- School of Engineering, University of Warwick, Coventry, UK.
2- Centre for Perinatal Research, School of Medicine, University of Nottingham, UK
3- Nottingham Centre for Neonatal Care, Queen's Medical Centre, Nottingham, UK.

*Corresponding Author: Declan G. Bates (d.bates@warwick.ac.uk)

**Funding:** DGB acknowledge funding from the UK Engineering and Physical Sciences Research Council (EP/W000490/1). SS acknowledges funding from UK Engineering and Physical Sciences Research Council (EP/Y003527/1) and for a Research Fellowship from the Royal Academy of Engineering (RF2122-21-258).

**Conflict of Interest:** The authors declare that they have no conflicts of interests.

## Abstract:

**Background:** Mechanical ventilation is life-saving for preterm infants with respiratory distress syndrome (RDS) but can also contribute to bronchopulmonary dysplasia (BPD) and long-term morbidity. Protective ventilation strategies are recommended, yet implementation in neonatal intensive care units remains inconsistent, and high-risk infants continue to be exposed to injurious ventilator settings.

**Objective:** To develop and validate a cohort of neonatal digital twins, based on mechanistic models of cardiopulmonary physiology calibrated to individual patient data, as a tool for simulating and optimising protective ventilation strategies.

**Methods:** A high-fidelity computational simulator of human cardiopulmonary physiology was adapted to neonatal-specific parameters, including lung compliance, dead space, pulmonary vascular resistance, oxygen consumption, and fetal haemoglobin oxygen affinity. Digital twins were generated using data at 65 time points from 11 preterm neonates receiving volume-controlled ventilation. Model parameters were calibrated to minimise the error between simulated and observed $PaO_2$, $PaCO_2$, and peak inspiratory pressure. The ability to recapitulate the measured data was assessed with correlation analysis and Bland–Altman comparisons.

**Results:** Digital twins reproduced measured data with mean absolute percentage errors of 3.9% ($PaO_2$), 3.0% ($PaCO_2$), and 5.8% (PIP) across the cohort. Predictions for uncalibrated variables (pHa, $SaO_2$, mean and minimum airway pressure) also showed high accuracy, with errors <5%. Strong correlations (R > 0.9) and narrow limits of agreement were observed across all patients and time points.

**Conclusions:** This study demonstrates, for the first time, the feasibility of creating fully mechanistic digital twins of mechanically ventilated neonates with RDS. The twins accurately captured patient-specific gas exchange and respiratory mechanics, supporting their potential as a platform for conducting virtual clinical trials and for the design of individualized, lung-protective ventilation strategies.

## Introduction:

Mechanical ventilation is an essential component of neonatal intensive care, particularly for very preterm infants with respiratory distress syndrome (RDS). Advances such as antenatal corticosteroids, surfactant therapy, and non-invasive respiratory support have improved outcomes [1], yet a substantial proportion of extremely preterm neonates still require invasive mechanical ventilation. While lifesaving, mechanical ventilation is also a major driver of bronchopulmonary dysplasia (BPD) and long-term neurodevelopmental impairment [1], [2]. Rates of moderate-to-severe BPD remain as high as 30–40% among infants born before 28 weeks' gestation, and survival without morbidity has not significantly improved in recent decades [2]. In low- and middle-income countries, mortality among ventilated neonates with RDS remains particularly high, ranging from 40% to 60%, underscoring the urgent global need for safer ventilation strategies [3].

To try to mitigate ventilator-induced lung injury (VILI), neonatal ventilation strategies have increasingly focused on lung-protection. Strong evidence supports volume-controlled ventilation (VCV), which reduces death or BPD, pneumothorax, hypocapnia, and severe intraventricular haemorrhage compared with pressure-limited ventilation [4], [5], [6], [7]. These benefits extend to infants with evolving BPD and are under evaluation in very-low-birthweight cohorts [8]. Permissive hypercapnia, tested in multiple randomized trials, has been shown to be feasible and safe, with recent evidence suggesting it can shorten ventilation duration and potentially reduce lung injury [2], [9]. High-frequency oscillatory ventilation (HFOV), once hypothesized to be highly protective, has not consistently shown superiority over conventional ventilation in large meta-analyses [10], [11], [12]. However, refinements such as HFOV with volume guarantee (HFOV+VG) may offer more stable $CO_2$ control and reduce fluctuations in tidal volume [13], [14].

Despite guideline recommendations that favour protective approaches, including early non-invasive support, VCV when intubated, and careful targeting of oxygen and carbon dioxide levels [1], [15], [16], clinical practice remains heterogeneous. Many high-risk infants continue to be managed with traditional pressure-controlled modes, and adherence to strict oxygen and $CO_2$ targets is difficult to maintain in the busy neonatal intensive care unit. Clinician workload has been directly linked to outcomes, and frequent life-threatening errors have been documented in intensive care environments, further complicating the challenge of delivering consistently protective ventilation [17], [18], [19].

These limitations highlight the need for new tools to facilitate research into improved neonatal ventilation strategies. Digital twins, mechanistic computational models of neonatal physiology matched to individual patient data, could allow for virtual clinical trials to explore novel ventilation strategies *in silico*, potentially leading to more individualized, lung-protective care. In this paper, we present and validate the first digital twins of neonatal patients undergoing invasive mechanical ventilation.

## Materials & Methods

*Cardiopulmonary simulator and adaptation to neonatal physiology:*

The core modules of the computational simulator used in this study have been developed over the past several years [20]. The model has been validated in several previous studies into new therapeutic invasive and non-invasive interventions for paediatric and adult patients [21], [22], [23]. The model represents multiple interacting organ systems (cardio-pulmonary-vascular). Multiple alveolar compartments (up to several hundreds), multicompartmental gas-exchange, non-linear viscoelastic alveolar compliance, interdependent blood-gas solubility and haemoglobin behaviour, and heterogeneous distributions of pulmonary ventilation and perfusion are all modelled explicitly.

Clinical observations, experimental data, and well accepted physiological relationships inform the algebraic equations underlying the model components. These equations are solved in series in an iterative manner, so that solving one equation at current time instant determines the values of the independent variables in the next equation. At the end of each iteration, the results of the solution of the final equations determine the independent variables of the first equation for the next iteration. The model simulates all relevant aspects of pulmonary dynamics and gas exchange – i.e. the transport of air from mouth to airway and alveoli (through the endotracheal tube), the gas exchange between alveoli and their corresponding capillaries, and the gas exchange between blood and peripheral tissue compartment. The model includes series deadspace (i.e. conducting airways where there is no gas exchange) to represent the tracheal tube, trachea, bronchi and non-respiratory bronchioles. The lung model incorporates multiple independently configurable alveolar compartments, implemented in parallel. Multiple alveolar compartments allow the model to simulate alveolar deadspace and venous admixture accurately. Figure 1 shows a simplified, diagrammatic representation of the model. For a full description of the original adult and paediatric models and their underlying mathematical principles the reader is referred to [21].

The extremely small volumes of neonates' lungs, as well as their large respiratory and vascular resistances make simulating the neonatal respiratory system challenging. Physiological features such as lung volume, cardiac output, oxygen consumption and airway resistance are weight-dependent in neonates, and some parameters such as pulmonary vascular resistance are highly variable during the first hours of life. To account for this, the cardiac output (CO) and the volume of functional residual capacity ($V_{frc}$) are estimated in the model using the following equations [24], [25]:

$$\begin{cases} CO = 265 \times weight & weight < 1.5 \ (kg) \\ CO = 253 \times weight & 1.5 < weight < 2.5 \ (kg) \\ CO = 241 \times weight & weight > 2.5 \ (kg) \end{cases} \quad (mL\ min^{-1})$$

$$V_{frc} = (20.7 \times weight) - 6.3 \qquad (mL)$$

The total airway resistance is notably higher in neonatal patients than in adults and decreases as they grow older. This resistance is distributed between the main airway and 50 parallel alveolar compartments in the model. Every alveolar compartment also has two resistances placed in series, namely the alveolar inlet resistance and the upper bronchial resistance. The pulmonary vascular resistance (PVR) is very large during foetal life, and drops sharply during the first 24 hours of life as gas exchange is the primary function of the postnatal lung [26], [27].

Both anatomical and alveolar deadspaces have been shown to increase with decreasing weight and gestation. Moreover, newborn babies have larger total anatomical dead space ($VD_{anat}$) per kg of body weight in comparison with adults, due to the larger head to body mass ratio. The value of $VD_{anat}$ is 3-6 mL kg$^{-1}$ in preterm infants weighing around 1 kg [28], [29]. Newborn babies have a higher haemoglobin concentration compared to adults, with a minimum value of about 120 g L$^{-1}$ for preterm infants. This value begins to fall within the first week after delivery [30], [31]. Oxygen consumption in neonates is also more than twice that of adults on a per kg basis. There is a large rise in metabolic rate in the first 24 hours for normal term babies; however, the rate of increase is slower in those born prematurely. Preterm babies also have a lower oxygen consumption rate compared with term babies. The optimized value for oxygen consumption in the model is selected from a range of 4-10 mL kg$^{-1}$ min$^{-1}$ [32], [33]. Moreover, fetal haemoglobin has a higher affinity for oxygen than adult haemoglobin due to its reduced interaction with 2,3-bisphosphoglycerate, resulting in a leftward shift of the oxygen dissociation curve [34], [35]. This physiological feature enhances placental oxygen transfer but requires accurate representation in neonatal respiratory models. For this reason, the oxygen dissociation curve used in the adult model was revisited and adapted to account for HbF-dominant blood, following the approach described in [36].

In the model, each single viscoelastic alveolar compartment is characterized by three main individually configurable parameters. $P_{ext}$ is a lumped parameter representing the extrinsic net pressure generated by the sum of the effects of factors outside each alveolus that act to distend/compress that alveolus, including the outward pulling force of the chest wall (positive effect) and the compressive effect of interstitial fluid in the alveolar wall (negative effect). A negative value of $P_{ext}$ indicates a scenario where there is compression from outside the alveolus causing collapse. $K_{stiff}$ represents the inherent stiffness of the alveolar compartment and is modelled in the adult simulator by $\frac{10^S}{20}$ where $10^S$ is a coefficient and equals 1 (i.e. S=0) for a healthy lung. The equation is changed to $K_{stiff} = \frac{10^S}{0.08}$ to characterize the stiffer nature of infant lungs at baseline. Finally, TOP is the alveolar compartment threshold opening pressure. The average threshold opening pressure of all the compartments ($TOP_{mean}$) is estimated to be around 20 cmH$_2$O in adults [37]. However, specific data on mean opening pressures in preterm neonates are lacking. Evidence from some studies suggests that threshold opening pressures in this population are generally lower, with the majority of infants requiring less than 20 cmH$_2$O. In paediatrics, airway opening has been shown to be visible at 15 cmH$_2$O [38]. In our model, we therefore adopted a shifted distribution of *TOP* values based on [37], adjusted to yield a reduced mean value of 10 cmH$_2$O. Selection of varying values for these three parameters across each of the multiple alveolar compartments in the model allows for a highly detailed representation of the heterogeneous nature of diseased lungs.

## *Patient data:*

Data were collected from 11 neonatal patients with RDS over a total of 65 different time points from the neonatal intensive care unit of the Queen's Medical Centre, Nottingham University Hospital, for the purposes of building the digital twins as part of the NeoPredict Study (East Midlands NHS Ethics Committee 18/EM/0033). Patients were under volume-controlled ventilation, and an endotracheal tube (3 mm internal diameter) was used for all patients. The relevant patient and mechanical ventilation data are listed in Table 1.

## *Construction of the digital twins:*

Digital twins were generated using the above described high-fidelity computational model of the human cardiopulmonary system. For each of the 65 time points included in the study, model parameters were calibrated using global optimization techniques to maximise agreement between simulated and observed measurements of partial pressures of oxygen and carbon dioxide in arterial blood (PaO$_2$, PaCO$_2$), and peak inspiratory pressure (PIP) during VCV. Optimization of the model to patient data was performed using the 'Blythe' high performance computing cluster provided by the University of Warwick, running MATLAB (2024b) and utilizing the global optimization and parallel computing toolboxes.

The model was calibrated against the individual patient data on arterial blood gas contents, airway pressures and ventilator settings for each patient in the dataset using an optimization approach. The model parameters (*x*) that were used in the optimization include two key alveolar features mentioned previously ($P_{ext}$, $k_{stiff}$) for each of the 50 alveolar compartments, as well as values for respiratory quotient (*RQ*), total oxygen consumption (*VO$_2$*), haemoglobin (*Hb*) and pulmonary vascular resistance. The optimization problem is formulated to find the configuration of model parameters (*x*) that minimize the difference between the model outputs (for a given set of ventilator settings) and the patient data. This error is captured by a cost function J given below:

$$\min_x J = \sqrt{\sum_{i=1}^{3} \frac{\hat{Y}_i - Y_i}{Y_i}}$$

Where

$$Y = [PaO_2, PaCO_2, PIP]$$

$Y$ is a vector of data values and $\hat{Y}$ is the model estimated values. Table 2 presents a summary of the parameters included in (*x*), with their dimensions and allowable range of variation.

# Results

*Outputs of the digital twins vs. patient data:*

Across the 65 neonatal digital twins generated, the simulator demonstrated close agreement with measured arterial blood gas values and airway pressures. Correlation plots (Figure 2) showed strong linear relationships between model predictions and patient data, with Pearson correlation coefficients exceeding 0.9 for $PaO_2$, $PaCO_2$, and PIP. The mean absolute percentage errors were 3.9% for $PaO_2$, 3.0% for $PaCO_2$, and 5.8% for PIP, indicating a high degree of accuracy in the calibrated twins. Bland–Altman analysis (Figure 3) confirmed low bias and narrow limits of agreement, with >95% of data points falling within the predefined limits.

Importantly, the predictive accuracy of the digital twins extended to variables not directly included in the calibration cost function. For pHa, $SaO_2$, mean airway pressure (mPaw), and minimum airway pressure (Pmin), mean absolute percentage errors were 0.3%, 2.6%, 4.7%, and 4.2%, respectively (Figure 4). This supports the mechanistic validity of the model, demonstrating its ability to reproduce integrated cardiopulmonary dynamics beyond the calibration variables.

Performance was consistent across patients of differing weights and severity of RDS (Table 1), underscoring the adaptability of the modelling framework. The digital twins maintained predictive accuracy even at time points where patient $PaO_2$ and $PaCO_2$ values were near the extremes of the observed ranges, reflecting robustness across the clinical spectrum. These results highlight the potential of digital twins to provide reliable patient-specific predictions in neonatal intensive care, and to generate clinically relevant indices such as mechanical power and stress/strain that cannot be measured directly at the bedside.

# Discussion

We have presented the first results from the development of a detailed, high-fidelity computational simulator of neonatal cardiorespiratory physiology. In this study, the model was shown to accurately reproduce clinical data from 11 preterm neonates across 65 time points, capturing arterial blood gas values and ventilatory pressures with high precision. Importantly, the mechanistic structure of the model ensured that predictive accuracy could be checked to extend beyond the calibrated variables i.e. simulated outputs for parameters not included in the cost function also remained within clinically acceptable error margins. This underlines the potential of mechanistic digital twins not only to replicate observed data but also to provide in silico predictions of clinically relevant measures that are otherwise unavailable at the bedside. For example, the twins can estimate indices such as mechanical power [39], which has been associated with VILI in both experimental and clinical studies in adult patients, yet is not routinely calculated in neonatal practice. Furthermore, the simulator enables the computation of VILI markers such as lung stress and strain [40], measures that directly reflect the mechanical load imposed on the lung parenchyma during. These indices are difficult to measure directly in critically ill neonates, but could provide valuable insights into the pathophysiological processes that drive lung injury arising from mechanical ventilation.

Despite clear evidence supporting lung-protective ventilation strategies, their implementation in neonatal intensive care units remains inconsistent. The European Consensus Guidelines on RDS emphasize maintaining narrow ranges for gas exchange and the early use of non-invasive support [1], but these targets are difficult to sustain in practice, and adherence to oxygen and carbon dioxide targets is often suboptimal [17], [18], [19]. Clinical workload and the dynamic instability of preterm infants contribute to deviations, with studies showing frequent life-threatening errors and substantial variability in care [2], [3]. Consequently, rates of BPD remain high, and survival without major morbidity has plateaued over the past decades. Mechanistic digital twins of the neonatal cardiopulmonary system offer a potential means to overcome these limitations. By mechanistically reproducing individual patient physiology, such simulators could allow for prediction of responses to ventilator adjustments in silico, before interventions are applied at the bedside. Previous work in paediatrics and adults has demonstrated that high-fidelity simulators can replicate patient-specific gas exchange and hemodynamic responses, and can be used to evaluate recruitment manoeuvres and optimize ventilator strategies. Extending these approaches to neonates creates the opportunity for virtual clinical trials in high-risk populations, where direct experimental studies are constrained by ethical and logistical barriers. Digital twins could thereby support decision-making, facilitate the development of closed-loop ventilation systems, and ultimately help deliver more consistently protective ventilation in the neonatal intensive care unit. Our findings suggest that neonatal digital twins could bridge the gap between limited bedside monitoring and the complex physiological information required to guide truly protective ventilation strategies.

This study has several limitations. First, the data were obtained from a single institution; although the severity of RDS and clinical outcomes were comparable to other reported cohorts, the broader generalizability of the findings remains to be established. Currently, the simulator can be calibrated only to venous or arterial blood gas measurements. This is possible because arterial and venous samples represent fully oxygenated and deoxygenated blood, respectively, thereby providing a clear point of entry for integration into the model. In clinical practice, however, arterial or venous sampling in neonates is often avoided for safety reasons, and capillary blood gases are generally preferred. The use of capillary samples poses additional challenges for model calibration, as gas exchange at the tissue level may be incomplete, making these measurements less directly comparable within the model framework. To reduce potential confounding, the model was configured to represent fully sedated patients under mechanical ventilation, and consequently, autonomic reflex pathways were not incorporated. Furthermore, the model does not account for the influence of inflammatory mediators commonly present in RDS, as these are difficult to isolate and quantify in clinical settings. Due to limited data for parameterization and uncertainties regarding certain aspects of lung physiology, the model may not represent all biophysical mechanisms relevant to VILI, such as airway closure caused by liquid bridge formation, etc. Nevertheless, the model reproduced all features of the available clinical data with high fidelity.

## Conclusions

This study has presented the first results from a high-fidelity computational simulator of neonatal cardiorespiratory physiology, marking an important step toward the application of digital twin technologies in this vulnerable patient group. The model achieved close agreement with clinical data, underscoring its ability to capture essential physiological dynamics. This approach provides a novel and useful platform for exploring treatment strategies that cannot be systematically tested at the bedside, with direct implications for improving ventilator management, refining clinical guidelines, and supporting the development of advanced closed-loop ventilation modes tailored to neonatal physiology.

Table 1. Patient characteristics and ventilator parameters presented as mean ± standard deviation

| Characteristic | Patients Involved (N=11) Timpe Points ($N_{TP}$=65) |
|---|---|
| Age (weeks) | 25.4±1.0 |
| Weight (g) | 785.6±138.8 |
| Oxygenation Index | 4.0±1.6 |
| $PaCO_2$ (kPa) | 5.9±1.3 |
| $PaO_2$ (kPa) | 8.0±2.3 |
| $SaO_2$ (%) | 93.1±2.3 |
| pHa | 7.24±0.1 |
| PEEP ($cmH_2O$) | 5.6±0.7 |
| Respiratory Rate (bpm) | 48.4±7.7 |
| Tidal Volume (ml) | 3.6±0.7 |
| I:E Ratio | 0.26±0.05 |
| $FiO_2$ | 0.27±0.09 |
| PF Ratio (mmHg) | 237.1±53.5 |

Table 2. List of the parameters varied by the optimization algorithm in order to calibrate the model to patient data, with their dimensions and allowable range of variation

| Parameter (x) | Size | Range |
|---|---|---|
| $P_{ext}$ | 50 | [-25,25] |
| $k_{stiff}$ | 50 | [-1.4,1.8] |
| RQ | 1 | [0.7,1] |
| $VO_2$ (mL $kg^{-1}$ $min^{-1}$) | 1 | [4,10] |
| Hb (g $L^{-1}$) | 1 | [120,200] |
| PVR (dyne s $cm^{-5}$) | 1 | [100,300] |

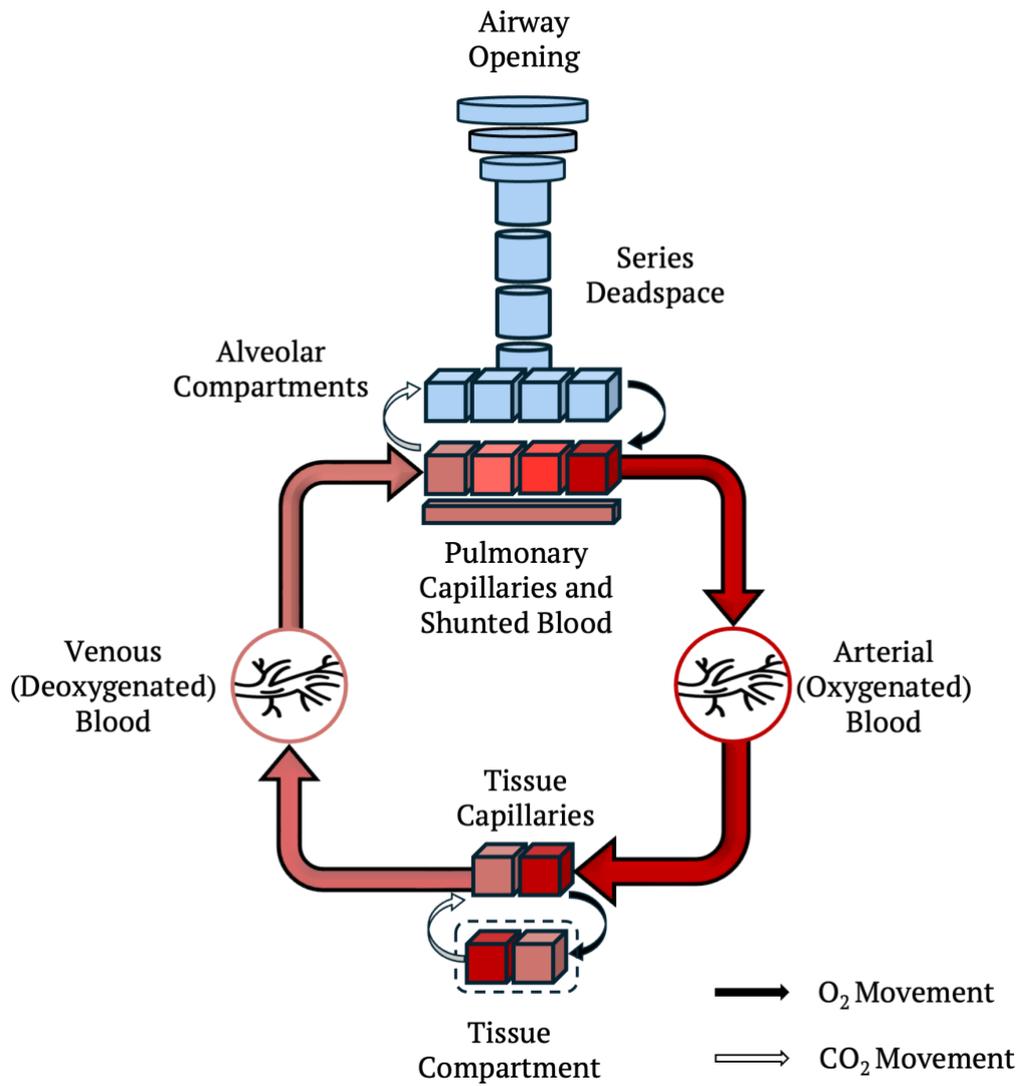

*Figure 1. Diagrammatic representation of the simulator*

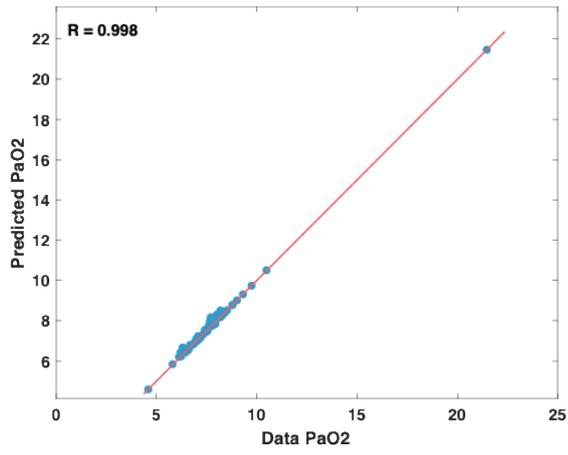
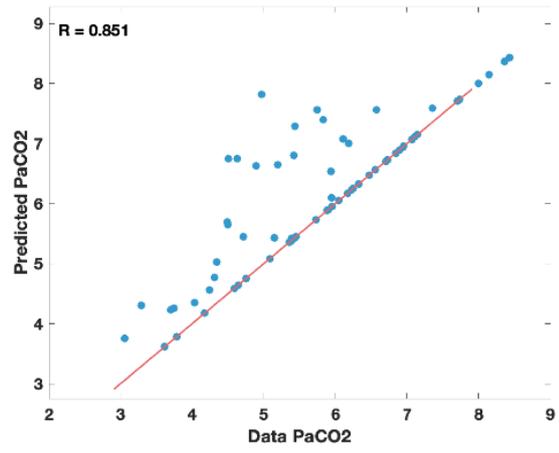
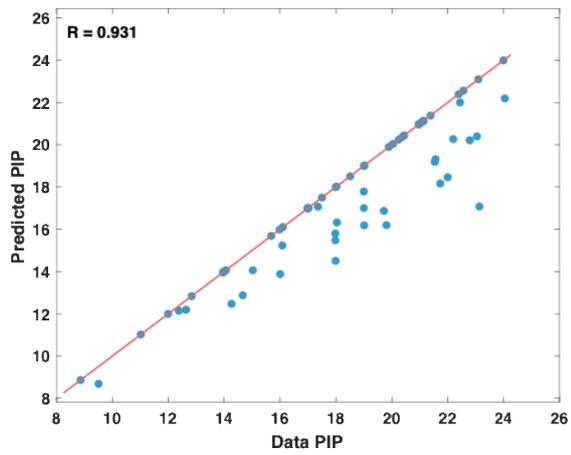

*Figure 2. Correlation plot showing patient data compared to digital twin outputs for $PaO_2$, $PaCO_2$ and PIP, R shows the Pearson correlation coefficient.*

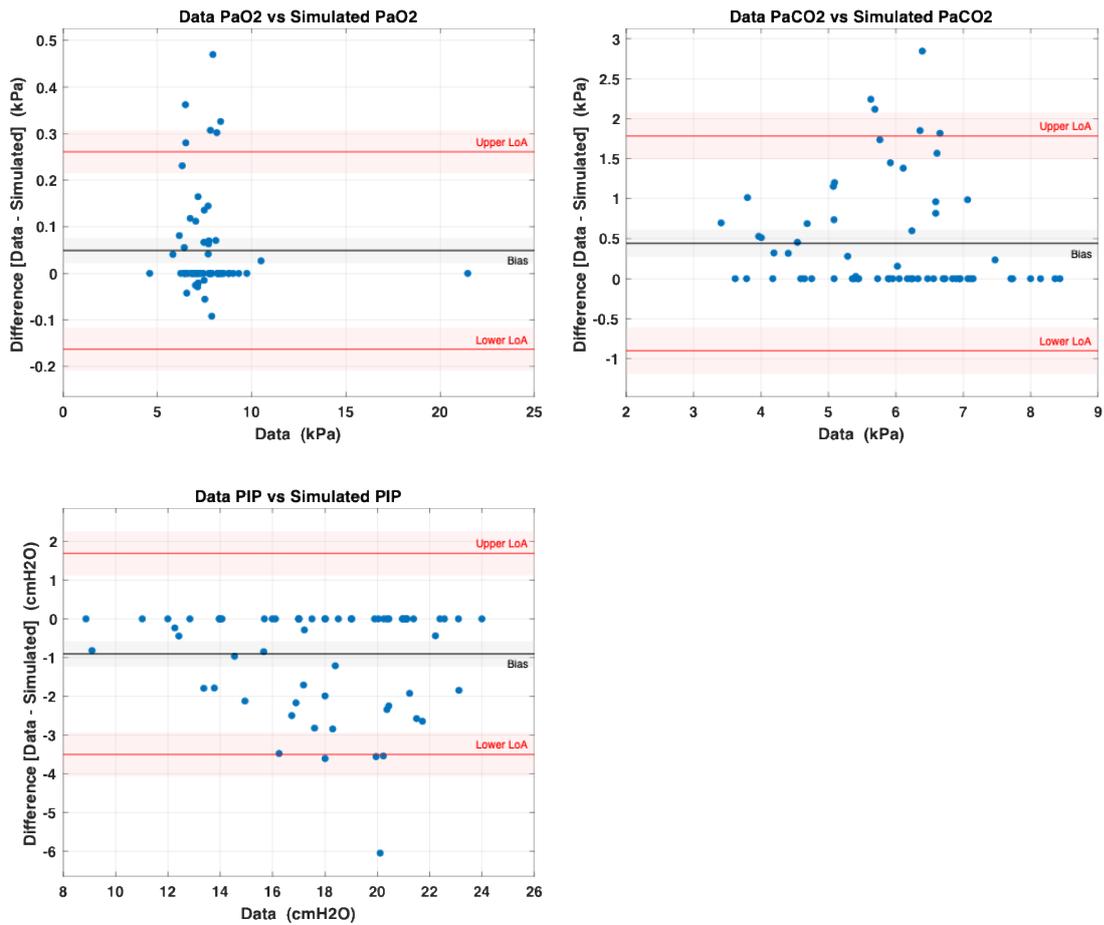

*Figure 3. Bland–Altman plot comparing measurements from data and simulated values. The solid black line indicates the mean bias between methods, and the red lines show the 95% limits of agreement (mean bias ± 1.96 SD). Shaded regions denote the 95% confidence interval for the bias and limits of agreement. Each point represents one paired measurement.*

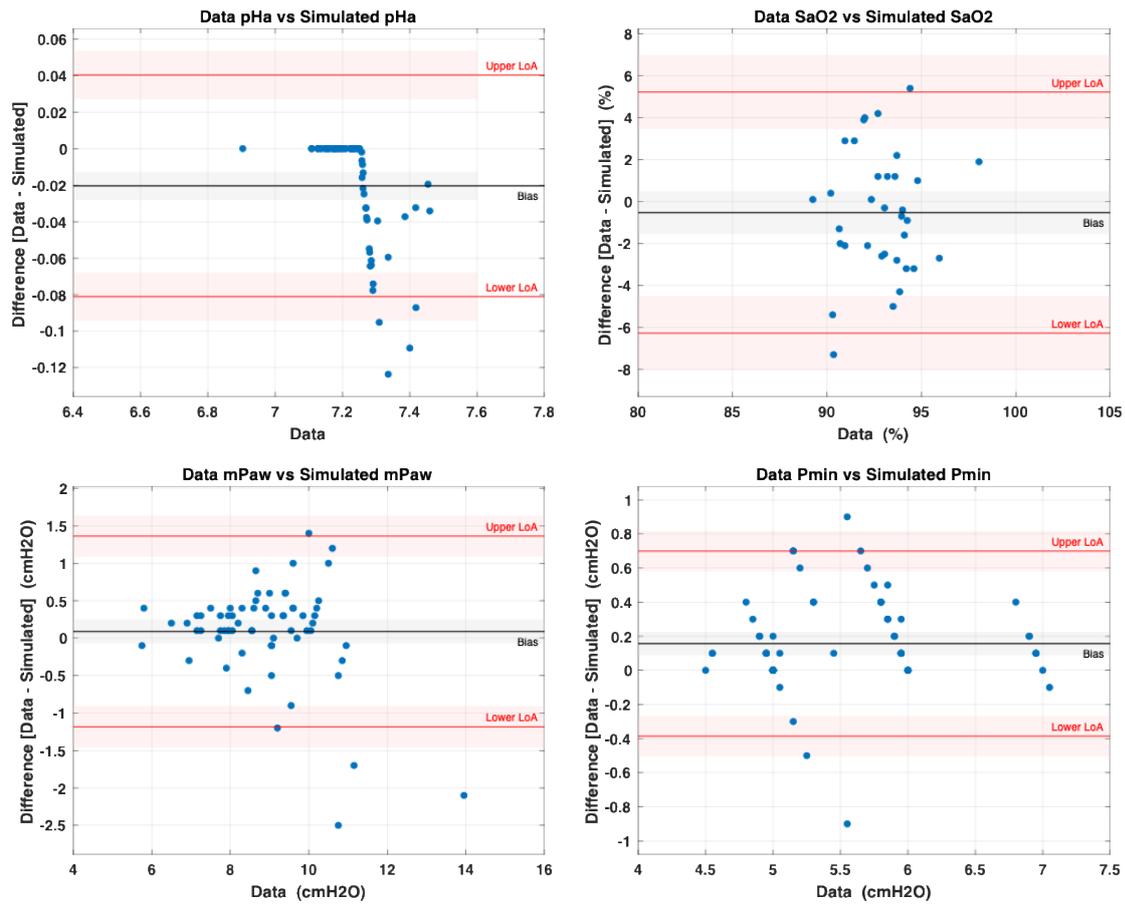

*Figure 4. Bland–Altman plot comparing measurements from data and simulated values. The solid black line indicates the mean bias between methods, and the red lines show the 95% limits of agreement (mean bias ± 1.96 SD). Shaded regions denote the 95% confidence interval for the bias and limits of agreement. Each point represents one paired measurement.*

# References

[1] D. G. Sweet *et al.*, "European Consensus Guidelines on the Management of Respiratory Distress Syndrome: 2022 Update," *Neonatology*, vol. 120, no. 1, pp. 3–23, Mar. 2023, doi: 10.1159/000528914.

[2] C. P. Travers *et al.*, "Late Permissive Hypercapnia for Mechanically Ventilated Preterm Infants: A Randomized Trial," *Pediatr Pulmonol*, vol. 60, no. 6, Jun. 2025, doi: 10.1002/ppul.71165.

[3] A. Ganguly, A. Makkar, and K. Sekar, "Volume Targeted Ventilation and High Frequency Ventilation as the Primary Modes of Respiratory Support for ELBW Babies: What Does the Evidence Say?," Feb. 07, 2020, *Frontiers Media S.A.* doi: 10.3389/fped.2020.00027.

[4] K. I. Wheeler, C. Klingenberg, C. J. Morley, and P. G. Davis, "Volume-targeted versus pressure-limited ventilation for preterm infants: A systematic review and meta-analysis," Sep. 2011. doi: 10.1159/000326080.

[5] N. Duman, F. Tuzun, S. Sutcuoglu, C. D. Yesilirmak, A. Kumral, and H. Ozkan, "Impact of volume guarantee on synchronized ventilation in preterm infants: A randomized controlled trial," *Intensive Care Med*, vol. 38, no. 8, pp. 1358–1364, Aug. 2012, doi: 10.1007/s00134-012-2601-5.

[6] M. Cuevas Guaman, J. Hagan, D. Sabic, D. M. Tillman, and C. J. Fernandes, "Volume-guarantee vs. pressure-limited ventilation in evolving bronchopulmonary dysplasia," *Front Pediatr*, vol. 10, Dec. 2022, doi: 10.3389/fped.2022.952376.

[7] M. Keszler and M. K. Abubakar, "Volume-targeted ventilation," *Semin Perinatol*, vol. 48, no. 2, Mar. 2024, doi: 10.1016/j.semperi.2024.151886.

[8] J. Tang *et al.*, "Volume-targeted ventilation vs pressure-controlled ventilation for very low birthweight infants: a protocol of a randomized controlled trial," *Trials*, vol. 24, no. 1, Dec. 2023, doi: 10.1186/s13063-023-07564-x.

[9] G. Mariani, J. Cifuentes, and W. A. Carlo, "Randomized Trial of Permissive Hypercapnia in Preterm Infants," 1999. [Online]. Available: http://publications.aap.org/pediatrics/article-pdf/104/5/1082/840472/1082.pdf

[10] F. Cools, D. J. Henderson-Smart, M. Offringa, and L. M. Askie, "Elective high frequency oscillatory ventilation versus conventional ventilation for acute pulmonary dysfunction in preterm infants," *Cochrane Database of Systematic Reviews*, no. 3, Jan. 2015, doi: 10.1002/14651858.CD000104.pub4.

[11] A. Rrt *et al.*, "Articles Elective high-frequency oscillatory versus conventional ventilation in preterm infants: a systematic review and meta-analysis of individual patients' data," *Lancet*, vol. 375, pp. 2082–91, 2010, doi: 10.1016/S0140.

[12] T. Bhuta and D. J. Henderson-Smart, "Rescue high frequency oscillatory ventilation versus conventional ventilation for pulmonary dysfunction in preterm infants," *Cochrane Database of Systematic Reviews*, Apr. 1998, doi: 10.1002/14651858.cd000438.

[13] B. Iscan, N. Duman, F. Tuzun, A. Kumral, and H. Ozkan, "Impact of Volume Guarantee on High-Frequency Oscillatory Ventilation in Preterm Infants: A Randomized Crossover Clinical Trial," *Neonatology*, vol. 108, no. 4, pp. 277–282, Oct. 2015, doi: 10.1159/000437204.

[14] M. Tana *et al.*, "Effects of High-Frequency Oscillatory Ventilation With Volume Guarantee During Surfactant Treatment in Extremely Low Gestational Age Newborns With Respiratory Distress Syndrome: An Observational Study," *Front Pediatr*, vol. 9, Mar. 2022, doi: 10.3389/fped.2021.804807.

[15] U. H. Thome and N. Ambalavanan, "Permissive hypercapnia to decrease lung injury in ventilated preterm neonates," *Semin Fetal Neonatal Med*, vol. 14, no. 1, pp. 21–27, Feb. 2009, doi: 10.1016/j.siny.2008.08.005.

[16] G. Emeriaud *et al.*, "Executive Summary of the Second International Guidelines for the Diagnosis and Management of Pediatric Acute Respiratory Distress Syndrome (PALICC-2)," *Pediatric Critical Care Medicine*, vol. 24, no. 2, pp. 143–168, Feb. 2023, doi: 10.1097/PCC.0000000000003147.

[17] "Outcomes of Two Trials of Oxygen-Saturation Targets in Preterm Infants," *New England Journal of Medicine*, vol. 374, no. 8, pp. 749–760, Feb. 2016, doi: 10.1056/NEJMoa1514212.

[18] J. I. Hagadorn *et al.*, "Achieved Versus Intended Pulse Oximeter Saturation in Infants Born Less Than 28 Weeks' Gestation: The AVIOx Study," *Pediatrics*, vol. 118, no. 4, pp. 1574–1582, Oct. 2006, doi: 10.1542/peds.2005-0413.

[19] H. A. van Zanten, R. N. G. B. Tan, A. van den Hoogen, E. Lopriore, and A. B. te Pas, "Compliance in oxygen saturation targeting in preterm infants: a systematic review.," *Eur J Pediatr*, vol. 174, no. 12, pp. 1561–72, Dec. 2015, doi: 10.1007/s00431-015-2643-0.

[20] J. G. Hardman, N. M. Bedforth, A. B. Ahmed, R. P. Mahajan, and A. R. Aitkenhead, "A physiology simulator: validation of its respiratory components and its ability to predict the patient's response to changes in mechanical ventilation," *Br J Anaesth*, vol. 81, no. 3, pp. 327–332, Sep. 1998, doi: 10.1093/bja/81.3.327.

[21] S. Saffaran, A. Das, J. G. Hardman, N. Yehya, and D. G. Bates, "High-fidelity computational simulation to refine strategies for lung-protective ventilation in paediatric acute respiratory distress syndrome," *Intensive Care Med*, vol. 1, pp. 10–12, 2019, doi: 10.1007/s00134-019-05559-4.

[22] L. Weaver *et al.*, "High risk of patient self-inflicted lung injury in COVID-19 with frequently encountered spontaneous breathing patterns: a computational modelling study," *Ann Intensive Care*, vol. 11, no. 1, p. 109, Dec. 2021, doi: 10.1186/s13613-021-00904-7.

[23] L. Weaver *et al.*, "Digital Twins of Acute Hypoxemic Respiratory Failure Patients Suggest a Mechanistic Basis for Success and Failure of Noninvasive Ventilation," *Crit Care Med*, vol. 52, no. 9, pp. e473–e484, Sep. 2024, doi: 10.1097/CCM.0000000000006337.

[24] T. Gerhardt, L. Reifenberg, D. Hehre, R. Feller, and E. Bancalari, "Functional Residual Capacity in Normal Neonates and Children up to 5 Years of Age Determined by a N2 Washout Method," *Pediatr Res*, vol. 20, no. 7, pp. 668–671, Jul. 1986, doi: 10.1203/00006450-198607000-00018.

[25] F. J. Walther, B. Siassi, N. A. Ramadan, A. K. Ananda, and P. Y. Wu, "Pulsed Doppler determinations of cardiac output in neonates: normal standards for clinical use.," *Pediatrics*, vol. 76, no. 5, pp. 829–33, Nov. 1985.

[26] A. M. RUDOLPH, "The Changes in the Circulation After Birth," *Circulation*, vol. 41, no. 2, pp. 343–359, Feb. 1970, doi: 10.1161/01.CIR.41.2.343.

[27] V. C. Baum, K. Yuki, and D. G. de Souza, "Cardiovascular Physiology," in *Smith's Anesthesia for Infants and Children*, Elsevier, 2017, pp. 73-107.e6. doi: 10.1016/B978-0-323-34125-7.00004-8.

[28] T. Dassios, P. Dixon, A. Hickey, S. Fouzas, and A. Greenough, "Physiological and anatomical dead space in mechanically ventilated newborn infants," *Pediatr Pulmonol*, vol. 53, no. 1, pp. 57–63, Jan. 2018, doi: 10.1002/ppul.23918.

[29] A. H. Numa and C. J. Newth, "Anatomic dead space in infants and children," *J Appl Physiol*, vol. 80, no. 5, pp. 1485–1489, May 1996, doi: 10.1152/jappl.1996.80.5.1485.

[30] E. H. Kates and J. S. Kates, "Anemia and Polycythemia in the Newborn," *Pediatr Rev*, vol. 28, no. 1, pp. 33–34, Jan. 2007, doi: 10.1542/pir.28-1-33.

[31] O. Linderkamp, E. P. Zilow, and G. Zilow, "[The critical hemoglobin value in newborn infants, infants and children].," *Beitrage zur Infusionstherapie = Contributions to infusion therapy*, vol. 30, pp. 235–46; discussion 247-64, 1992.

[32] J. W. Scopes and I. Ahmed, "Minimal rates of oxygen consumption in sick and premature newborn infants.," *Arch Dis Child*, vol. 41, no. 218, pp. 407–16, Aug. 1966, doi: 10.1136/adc.41.218.407.

[33] J. Roze, J. Liet, V. Gournay, T. Debillon, and C. Gaultier, "Oxygen cost of breathing and weaning process in newborn infants," *European Respiratory Journal*, vol. 10, no. 11, pp. 2583–2585, Nov. 1997, doi: 10.1183/09031936.97.10112583.

[34] T. Ulinder *et al.*, "Fetal haemoglobin and oxygen requirement in preterm infants: an observational study," *Arch Dis Child Fetal Neonatal Ed*, vol. 110, no. 3, pp. 285–290, May 2025, doi: 10.1136/archdischild-2024-327411.

[35] T. Dassios, K. Ali, T. Rossor, and A. Greenough, "Using the fetal oxyhaemoglobin dissociation curve to calculate the ventilation/perfusion ratio and right to left shunt in healthy newborn infants," *J Clin Monit Comput*, vol. 33, no. 3, pp. 545–546, Jun. 2019, doi: 10.1007/s10877-018-0168-6.

[36] W. Käsinger, R. Huch, and A. Huch, "*In vivo* oxygen dissociation curve for whole fetal blood: fitting the Adair equation and blood gas nomogram," *Scand J Clin Lab Invest*, vol. 41, no. 7, pp. 701–707, Jan. 1981, doi: 10.3109/00365518109090518.

[37] S. Crotti *et al.*, "Recruitment and derecruitment during acute respiratory failure: a clinical study.," *Am J Respir Crit Care Med*, vol. 164, no. 1, pp. 131–40, 2001, doi: 10.1164/ajrccm.164.1.2007011.

[38] L. Rodriguez Guerineau *et al.*, "Airway opening pressure maneuver to detect airway closure in mechanically ventilated pediatric patients.," *Front Pediatr*, vol. 12, p. 1310494, 2024, doi: 10.3389/fped.2024.1310494.

[39] L. Gattinoni *et al.*, "Ventilator-related causes of lung injury: the mechanical power," *Intensive Care Med*, vol. 42, no. 10, pp. 1567–1575, Oct. 2016, doi: 10.1007/s00134-016-4505-2.

[40] A. Protti, E. Votta, and L. Gattinoni, "Which is the most important strain in the pathogenesis of ventilator-induced lung injury," *Curr Opin Crit Care*, vol. 20, no. 1, pp. 33–38, 2014, doi: 10.1097/MCC.0000000000000047.